\documentstyle[12pt]{article}
\renewcommand{\theequation}{\arabic{section}.\arabic{equation}}

\catcode`\@=11
\@addtoreset{equation}{section}
\catcode`\@=12
\begin{document}
\thispagestyle{empty}
\baselineskip8mm
\title{\vspace{-2cm}
\bf Gravitons in one-loop quantum cosmology:
correspondence between covariant and non-covariant formalisms}
\author{Giampiero Esposito$^{1,2}$,\thanks{Electronic address:
esposito@na.infn.it} \ Alexander Yu. Kamenshchik,${^3}$\thanks
{Electronic address: grg@ibrae.msk.su}\\ \ Igor V. Mishakov$^{3}$ and
\ Giuseppe Pollifrone$^{1,4}$\thanks{Electronic address:
pollifrone@sci.uniroma1.it}}
\date{}
\maketitle
\hspace{-6mm}$^{1}${\em Istituto Nazionale di Fisica Nucleare,
Mostra d'Oltremare Padiglione 20, 80125 Napoli, Italy}\\ $^{2}${\em
Dipartimento di Scienze Fisiche, Mostra d'Oltremare Padiglione 19,
80125 Napoli, Italy}\\ $^{3}${\em Nuclear Safety Institute, Russian
Academy of Sciences, 52 Bolshaya Tulskaya, Moscow, 113191, Russia}\\
$^{4}${\em Dipartimento di Fisica, Universit\`a di Roma ``La
Sapienza", Piazzale Aldo Moro 2, 00185 Roma, Italy}

{\bf Abstract}. The discrepancy between the results of
covariant and non-covariant one-loop calculations for higher-spin
fields in quantum cosmology is analyzed.
A detailed mode-by-mode study of perturbative quantum gravity
about a flat Euclidean background bounded by two concentric
3-spheres, including non-physical degrees of freedom and
ghost modes, leads to one-loop amplitudes in agreement with
the covariant Schwinger-DeWitt method. This calculation provides
the generalization of a previous analysis of fermionic fields
and electromagnetic fields at one-loop about flat Euclidean
backgrounds admitting a well-defined 3+1 decomposition.

\section {INTRODUCTION}
\hspace{\parindent}
In recent papers devoted to one-loop calculations in quantum
gravity and quantum cosmology [1-11], discrepancies were found
between the covariant Schwinger-DeWitt
method [12], where the scaling factor
of one-loop amplitudes coincides with the $A_{2}$ coefficient
in the heat-kernel expansion, and zeta-function regularization
[13-14] with the corresponding mode-by-mode
analysis of quantized fields.

In Ref. [1] it was shown that the $\zeta(0)$ value for gravitons
calculated on the Riemannian de Sitter 4-sphere in terms
of physical degrees of freedom [15] disagrees with that obtained in
covariant formalism [16]. In Refs. [2-11] it was shown that the
results of analogous calculations for gauge fields on
manifolds with boundaries, as well as earlier results of Refs.
[17-18], are in disagreement with the results of covariant
calculations using the modified Schwinger-DeWitt formulae for
manifolds with boundaries [19].

In Ref. [20] the attempt of investigating the reason of the
discrepancy described above was made. Since for spin-${1\over 2}$
fields there is no gauge freedom, different results cannot
be due to inequivalent quantization techniques, but may have
an entirely geometrical origin. The hypothesis was then put
forward that the reason of this discrepancy consists in the
inappropriate use of 3+1 decomposition on the manifolds where such
a split is ill-defined (apparently the 3+1 decomposition is
ill-defined on the full 4-dimensional sphere or on the part of
4-sphere or 4-dimensional flat space bounded
by a single 3-sphere, since the vector
field matching the normal to the boundary is singular at the origin).
The $\zeta(0)$ value for massless spin-1/2 fields on the manifold
representing the part of flat Euclidean 4-dimensional space
bounded by two concentric 3-spheres (such a manifold admits a
well-defined 3+1 split) was calculated. It was shown that
the result coincides with the covariant one obtained
within the framework of the Schwinger-DeWitt
method [3,19].

On evaluating $\zeta(0)$ for vacuum Maxwell theory
subject to the Coulomb gauge {\it before} quantization,
the one-loop result is in
disagreement with the covariant one, even in the two-boundary
case, where one has a well-defined 3+1 split [21].
The possible reason is that,
on non-trivial backgrounds with boundaries,
contributions of ghosts and non-physical degrees of freedom
do not cancel each other.
In Ref. [21], continuing the study of the
quantization of the electromagnetic field on manifolds with
boundaries appearing in Ref. [22],
this suggestion was checked. It was shown that, if one deals
with non-physical degrees of freedom and ghost modes within
the Faddeev-Popov formalism, and if one studies a flat Euclidean
background bounded by two concentric 3-spheres, the one-loop
evaluation of quantum amplitudes agrees with the covariant result,
confirming in such a way our hypotheses. A technique for the
disentanglement of coupled gauge modes was also described and
applied for the first time.

Here, we calculate $\zeta(0)$ for gravitons on the part of flat
4-dimensional Euclidean space bounded by two concentric
3-spheres, taking into account the contributions of non-physical
modes and ghosts. We show that in this case, which is more
complicated from the technical point of view than the
electromagnetic analysis of Refs. [18,21-22], one finds again
agreement between the results of covariant
and non-covariant calculations.

In our calculations
we use the version of the generalized $\zeta$-function technique
elaborated in [8-10]. The main ideas are as follows.
Zeta-functions are traces of complex powers of elliptic,
self-adjoint, positive-definite differential operators. They
have an analytic continuation to the complex plane as meromorphic
functions with simple poles. Remarkably, they are regular at
the origin together with their first derivative, and this
enables one to define and compute the determinants of the
corresponding operators and one-loop quantum amplitudes.
With our notation, one writes $f_{n}(M^{2})$ for the function
occurring in the equation obeyed by the eigenvalues by virtue of
boundary conditions, and $d(n)$ for the degeneracy of the
eigenvalues. One then defines the function
\begin{equation}
I(M^{2},s) \equiv \sum_{n=n_{0}}^{\infty}d(n)n^{-2s}
\log f_{n}(M^{2}).
\end{equation}
Such a function has a unique analytic continuation to the whole
complex-$s$ plane as a meromorphic function, i.e.
\begin{equation}
I(M^{2},s)={I_{\rm pole}(M^{2})\over s}
+I^{R}(M^{2})+O(s).
\end{equation}
The $\zeta(0)$ value is then obtained as [8-10]
\begin {equation}
\zeta(0) = I_{\log} + I_{\rm pole}(\infty) - I_{\rm pole}(0),
\end{equation}
where $I_{\rm log}=I_{\rm log}^{R}$ is the coefficient of
$\log M$ from $I(M^{2},s)$ as $M \rightarrow \infty$, and
$I_{\rm pole}(M^{2})$ is the residue at $s=0$. Remarkably,
$I_{\rm log}$ and $I_{\rm pole}(\infty)$ are obtained from
the uniform asymptotic expansions of modified Bessel functions
as their order tends to $\infty$ and $M \rightarrow \infty$,
whereas $I_{\rm pole}(0)$ is obtained from the limiting
behaviour of such Bessel functions as $M \rightarrow 0$.

In Sec. II we write down
the equations for basis functions for physical and non-physical
degrees of freedom of the gravitational field
with the corresponding ghost modes, and we find
their solutions on the chosen background. In Sec. III we consider the
boundary conditions and compute all
contributions to the full $\zeta(0)$ value.
Results and concluding remarks are presented in Sec. IV.
The forms of the differential operators acting on metric
perturbations are given in the Appendix.

\section {EQUATIONS FOR BASIS FUNCTIONS AND THEIR SOLUTIONS}
\hspace {\parindent}
For the reasons described in the introduction, we study pure
gravity at one-loop about a flat Euclidean background with
two concentric 3-sphere boundaries. Our approach to
quantization follows the Faddeev-Popov formalism (cf. [22]).
Hence we deal with quantum amplitudes of the form
$$
Z[{\rm boundary \; data}]=\int_{C} \mu_{1}[g] \; \mu_{2}[\varphi]
\; {\rm exp}(-{\widetilde I}_{E}).
$$
With our notation, $C$ is the set of all Riemannian 4-geometries
matching the boundary data,
$\mu_{1}$ is a suitable measure on the space
of metrics, $\mu_{2}$ is a suitable measure for ghosts,
$\Phi_{\nu}$ is an arbitrary gauge-averaging functional, and the
total Euclidean action reads (in $c=1$ units)
\begin{eqnarray}
&&{\widetilde I}_{E}=I_{gh}+{1\over 16 \pi G}
\int_{M}{ }^{(4)}R \sqrt{{\rm det} \; g} \; d^{4}x
+{1\over 8 \pi G} \int_{\partial M}{\rm Tr} \; K
\sqrt{{\rm det} \; q} \; d^{3}x \nonumber \\
&&+ {1\over 16\pi G} \int_{M}
{1\over 2\alpha} \Phi_{\nu}\Phi^{\nu}
\sqrt{{\rm det} \; g} \; d^{4}x .
\end{eqnarray}
Of course, $K$ is the extrinsic-curvature tensor of the
boundary, $q$ is the induced 3-metric of $\partial M$,
and $\alpha$ is a positive dimensionless parameter.
The ghost action $I_{gh}$ depends on the specific form of
$\Phi_{\nu}$. Denoting by $h_{\mu \nu}$ the perturbation
around the background 4-metric $g_{\mu \nu}$, one thus finds
equations of motion of the kind
$$
\Box^{\Phi} h_{\mu \nu} = 0 ,
$$
where $\Box^{\Phi}$ is the 4-dimensional elliptic operator
corresponding to the form of $\Phi_{\nu}$ one is working
with. Here we choose the de Donder gauge-averaging functional
$$
\Phi_{\nu}^{DD} \equiv
\nabla^{\mu} \left(h_{\mu \nu}-\frac{1}{2} g_{\mu
\nu} {\hat h} \right),
$$
where $\nabla^{\mu}$ is covariant differentiation with respect
to $g_{\mu \nu}$, and ${\hat h} \equiv g^{\mu \nu} \; h_{\mu \nu}$.
The corresponding $\Box^{\Phi^{DD}}$ operator is the one obtained
by analytic continuation of the standard D'Alembert operator,
hereafter denoted by $\Box$.
The resulting eigenvalue equation is (see Appendix)
$$
\Box h_{\mu \nu} + \lambda h_{\mu \nu} = 0.
$$

Now we can make the 3+1 decomposition of our background 4-geometry and
expand $h_{00}, h_{0 i}$ and $h_{ij}$ in hyperspherical harmonics
as
\begin{equation}
h_{00}(x,\tau) = \sum_{n=1}^{\infty} a_{n}(\tau) Q^{(n)}(x),
\end{equation}
\begin{equation}
h_{0i}(x,\tau) = \sum_{n=2}^{\infty}
\biggr[b_{n}(\tau) \frac{\nabla_{i}
Q^{(n)}(x)}{(n^{2} - 1)} +
c_{n}(\tau) S_{i}^{(n)}(x)\biggr],
\end{equation}
\begin{eqnarray}
&&h_{ij}(x,\tau) = \sum_{n=3}^{\infty}
d_{n}(\tau) \left(\frac{\nabla_{i} \nabla_{j} Q^{(n)}(x)}
{(n^{2} - 1)} + {c_{ij}\over 3} Q^{(n)}(x)\right)
+\sum_{n=1}^{\infty}{e_{n}(\tau)\over 3} c_{ij} Q^{(n)}(x)
\nonumber \\
&&+\sum_{n=3}^{\infty}\biggr[f_{n}(\tau)
\Bigr(\nabla_{i} S_{j}^{(n)}(x) + \nabla_{j} S_{i}^{(n)}(x)\Bigr)
+k_{n}(\tau) G_{ij}^{(n)}(x)\biggr].
\end{eqnarray}
Here $Q^{(n)}(x), S_{i}^{(n)}(x)$  and $G_{ij}^{(n)}(x)$ are scalar,
transverse vector and transverse-traceless tensor hyperspherical
harmonics respectively, on a unit 3-sphere with metric $c_{ij}$.
Their properties are described in Refs. [17-18,23].

The insertion of the expansions (2.2)-(2.4) into Eq. (2.1)
leads to the following system of equations
(decoupled modes will be treated separately):
\begin{equation}
{\widehat A}_{n} a_{n}(\tau) + {\widehat B}_{n} b_{n}(\tau)
+ {\widehat C}_{n} e_{n}(\tau) = 0,
\end{equation}
\begin{equation}
{\widehat D}_{n} b_{n}(\tau) + {\widehat E}_{n} a_{n}(\tau)
+ {\widehat F}_{n} d_{n}(\tau) + {\widehat G}_{n} e_{n}(\tau)
= 0,
\end{equation}
\begin{equation}
{\widehat L}_{n} d_{n}(\tau) + {\widehat M}_{n} b_{n}(\tau) = 0,
\end{equation}
\begin{equation}
{\widehat N}_{n} e_{n}(\tau) + {\widehat P}_{n} b_{n}(\tau)
+ {\widehat Q}_{n} a_{n}(\tau) = 0,
\end{equation}
\begin{equation}
{\widehat H}_{n} c_{n}(\tau) + {\widehat K}_{n} f_{n}(\tau) = 0,
\end{equation}
\begin{equation}
{\widehat R}_{n} f_{n}(\tau)
+ {\widehat S}_{n} c_{n}(\tau) = 0,
\end{equation}
\begin{equation}
{\widehat T}_{n} k_{n}(\tau) = 0.
\end{equation}
Since our background is flat,
after setting $\alpha=1$ in (2.1) the operators appearing in
Eqs. (2.5)-(2.11) take the form (for all integer $n \geq 3$)
\[{\widehat A}_{n} \equiv
\frac{d^{2}}{d \tau^{2}} + \frac{3}{\tau} \frac{d}{d
\tau} - \frac{(n^{2} + 5)}{\tau^{2}} +  \lambda_{n},\]
\[{\widehat B}_{n} \equiv \frac{4}{\tau^{3}},\]
\[{\widehat C}_{n} \equiv \frac{2}{\tau^{4}},\]
\[{\widehat D}_{n} \equiv
\frac{d^{2}}{d \tau^{2}} + \frac{1}{\tau} \frac{d}{d
\tau} - \frac{(n^{2} + 4)}{\tau^{2}} +  \lambda_{n},\]
\[{\widehat E}_{n} \equiv \frac{2}{\tau} (n^{2} - 1),\]
\[{\widehat F}_{n} \equiv \frac{4}{3} \frac{(n^{2} - 4)}{\tau^{3}},\]
\[{\widehat G}_{n} \equiv -\frac{2}{3} \frac{(n^{2} - 1)}{\tau^{3}},\]
\[{\widehat H}_{n} \equiv  \frac{d^{2}}{d \tau^{2}}
+ \frac{1}{\tau} \frac{d}{d
\tau} - \frac{(n^{2} + 5)}{\tau^{2}} +  \lambda_{n},\]
\[{\widehat K}_{n} \equiv \frac{2}{\tau^{3}} (n^{2} - 4),\]
\[{\widehat L}_{n} \equiv \frac{d^{2}}{d \tau^{2}}
- \frac{1}{\tau} \frac{d}{d
\tau} - \frac{(n^{2} - 5)}{\tau^{2}} +  \lambda_{n},\]
\[{\widehat M}_{n} \equiv \frac{4}{\tau},\]
\[{\widehat N}_{n} \equiv \frac{d^{2}}{d \tau^{2}}
- \frac{1}{\tau} \frac{d}{d
\tau} - \frac{(n^{2} + 1)}{\tau^{2}} +  \lambda_{n},\]
\[{\widehat P}_{n} \equiv -\frac{4}{\tau},\]
\[{\widehat Q}_{n} \equiv 6,\]
\[{\widehat R}_{n} \equiv \frac{d^{2}}{d \tau^{2}}
- \frac{1}{\tau} \frac{d}{d
\tau} - \frac{(n^{2} - 4)}{\tau^{2}} +  \lambda_{n},\]
\[{\widehat S}_{n} \equiv \frac{2}{\tau},\]
\begin{equation}
{\widehat T}_{n} \equiv \frac{d^{2}}{d \tau^{2}}
- \frac{1}{\tau} \frac{d}{d
\tau} - \frac{(n^{2} - 1)}{\tau^{2}} +  \lambda_{n}.
\end{equation}
Inserting the operator ${\widehat T}_{n}$
from Eq. (2.12) into Eq. (2.11)
we can easily find the basis function describing the
transverse-traceless symmetric tensor harmonics which usually are
treated as physical degrees of freedom [1,15,17,23]
\begin{equation}
k_{n}(\tau) = \alpha_{1} \tau I_{n}(M \tau)
+ \alpha_{2} \tau K_{n}(M \tau),\ \ n=3,\ldots
\end{equation}
where $M = \sqrt{-\lambda}$ and $I$ and $K$
are modified Bessel functions.

However, the equations (2.5)-(2.8) for scalar-type gravitational
perturbations lead to a rather complicated entangled system as well
as Eqs. (2.9)-(2.10), describing vector perturbations. In
Ref. [21], where we have studied the analogous problem for the
electromagnetic field, a method was used to decouple a similar
entangled system for normal and longitudinal components of
the 4-vector potential. The idea
is that one can diagonalize a $2 \times 2$ operator matrix after
multiplying it by two functional matrices. In some cases
one can choose these functional matrices in such a way that the
transformed operator matrix is diagonal and the corresponding
differential equations for basis functions are decoupled.
However, in the case of scalar-type gravitational perturbations we
have a $4 \times 4$ operator matrix.
To diagonalize such a matrix it is
necessary to solve a system of 24 second-order algebraic
equations with 24 variables. This problem seems a rather cumbersome
one and we thus use another method.
For this purpose, we assume that the solution of
the system of equations (2.5)-(2.8) is some set of modified Bessel
functions with unknown index $\nu$. Let us look for a solution of this
system in the form
\begin{equation}
a_{n}(\tau) = \beta_{1} \frac{W_{\nu}(M\tau)}{\tau},
\end{equation}
\begin{equation}
b_{n}(\tau) = \beta_{2} W_{\nu}(M\tau),
\end{equation}
\begin{equation}
d_{n}(\tau) = \beta_{3} \tau W_{\nu}(M\tau),
\end{equation}
\begin{equation}
e_{n}(\tau) = \beta_{4} \tau W_{\nu}(M\tau).
\end{equation}
Here, $W_{\nu}$ is a linear combination of modified Bessel functions
$I_{\nu}$ and $K_{\nu}$ obeying the Bessel equation
\begin{equation}
\left(\frac{d^{2}}{d \tau^{2}} + \frac{1}{\tau} \frac{d}{d
\tau} - \frac{\nu^{2}}{\tau^{2}} -  M^{2}\right) W_{\nu}(M\tau) = 0.
\end{equation}
Now, inserting the functions (2.14)-(2.17) and the
corresponding operators from Eq. (2.12) into the system of equations
(2.5)-(2.8), and taking into account the Bessel equation (2.18), one
finds the following system of equations for $\beta_{1},
\beta_{2}, \beta_{3}$ and $\beta_{4}$:
\begin{eqnarray}
&&(\nu^{2} - n^{2} - 6)\beta_{1} + 4\beta_{2} + 2\beta_{4} = 0,
\nonumber \\
&&6(n^{2} - 1)\beta_{1} +3(\nu^{2} - n^{2} - 4)\beta_{2}
+ 4(n^{2} - 4)\beta_{3} - 2(n^{2} - 1)\beta_{4} =
0,\nonumber \\
&&4\beta_{2} + (\nu^{2} - n^{2} +4)\beta_{3} = 0,\nonumber \\
&&6\beta_{1} - 4\beta_{2} + (\nu^{2} - n^{2} -2)\beta_{4} = 0.
\end{eqnarray}
The condition for the existence of nontrivial solutions of the system
(2.19) is the vanishing of its determinant, i.e.
\begin{equation}
(\nu^{2} - n^{2})^{2}
\Bigr[(\nu^{2} - n^{2})^{2} - 8(\nu^{2} - n^{2})
- 16(n^{2} - 1)\Bigr] = 0.
\end{equation}
The roots of Eq. (2.20) are
\[\nu^{2} = n^{2},\ \nu^{2} = (n-2)^{2},\ \nu^{2} = (n+2)^{2}.\]
The positive values of $\nu$ provide the orders of modified
Bessel functions.
Now we can write down the $\beta$'s corresponding to
different values for $\nu$'s.
For $\nu = n$ one has
\begin{equation}
\beta_{4} = 3\beta_{1},\ \beta_{2} = \beta_{3} = 0,
\end{equation}
or
\begin{equation}
\beta_{1} = 0,\ \beta_{3} = -\beta_{2}, \beta_{4} = -2\beta_{2}.
\end{equation}
For $\nu = n - 2$ one has
\begin{equation}
\beta_{2} = (n + 1)\beta_{1},\ \beta_{3} = \frac{(n + 1)}{(n - 2)}
\beta_{1},\ \beta_{4} = -\beta_{1}.
\end{equation}
Last, for  $\nu = n + 2$ one has
\begin{equation}
\beta_{2} = -(n - 1)\beta_{1},\ \beta_{3} = \frac{(n - 1)}{(n + 2)}
\beta_{1},\ \beta_{4} = -\beta_{1}.
\end{equation}
Having the Eqs. (2.21)-(2.24) we can get the basis functions for
scalar-type gravitational perturbations (2.14)--(2.17)
\begin{eqnarray}
&a_{n}(\tau)& = \frac{1}{\tau}\Bigr(\gamma_{1}I_{n}(M\tau)
+\gamma_{3}I_{n-2}(M\tau) + \gamma_{4}I_{n+2}(M\tau) \nonumber
\\ && + \delta_{1}K_{n}(M\tau) +\delta_{3}K_{n-2}(M\tau) +
\delta_{4}K_{n+2}(M\tau) \Bigr),
\end{eqnarray}
\begin{eqnarray}
&b_{n}(\tau)& =\gamma_{2}I_{n}(M\tau) + (n + 1)
\gamma_{3}I_{n-2}(M\tau) \nonumber \\
&&- (n-1)\gamma_{4}I_{n+2}(M\tau)
+\delta_{2}K_{n}(M\tau) \nonumber \\
&&+(n+1)\delta_{3}K_{n-2}(M\tau) - (n-1)\delta_{4}K_{n+2}(M\tau),
\end{eqnarray}
\begin{eqnarray}
&d_{n}(\tau)& =\tau \left(-\gamma_{2}I_{n}(M\tau) +
\frac{(n+1)}{(n-2)}\gamma_{3}I_{n-2}(M\tau) \right. \nonumber \\
&&+ \frac{(n-1)}{(n+2)}\gamma_{4}I_{n+2}(M\tau)
-\delta_{2}K_{n}(M\tau) \nonumber \\
&& \left. +\frac{(n+1)}{(n-2)}\delta_{3}K_{n-2}(M\tau)
+ \frac{(n-1)}{(n+2)}\delta_{4}K_{n+2}(M\tau)\right)
\end{eqnarray}
\begin{eqnarray}
&e_{n}(\tau)& =\tau
\Bigr(3\gamma_{1}I_{n}(M\tau) - 2\gamma_{2}I_{n}(M\tau)
-\gamma_{3}I_{n-2}(M\tau) \nonumber \\
&&- \gamma_{4}I_{n+2}(M\tau)
+3\delta_{1}K_{n}(M\tau) - 2\delta_{2}K_{n}(M\tau) \nonumber \\
&&-\delta_{3}K_{n-2}(M\tau)
- \delta_{4}K_{n+2}(M\tau)\Bigr).
\end{eqnarray}

We can find the basis functions for vector-like gravitational
perturbations in a similar way. Let us suppose that
\begin{equation}
c_{n}(\tau) = \varepsilon_{1} W_{\nu}(M\tau)
\end{equation}
and
\begin{equation}
f_{n}(\tau) = \varepsilon_{2} \tau W_{\nu}(M\tau).
\end{equation}
Inserting (2.29)-(2.30) into Eqs.
(2.9)-(2.10) one has the system
\[(\nu^{2} - n^{2} - 5)\varepsilon_{1}
+ 2(n^{2} - 4)\varepsilon_{2} = 0,\]
\begin{equation}
2\varepsilon_{1} + (\nu^{2} - n^{2} + 3)\varepsilon_{2} = 0.
\end{equation}
The determinant of the system (2.31) is
\[(\nu^{2} - n^{2})^{2} - 2(\nu^{2} - n^{2}) - 4n^{2} + 1\]
and its positive roots are $n \pm 1$.
For $\nu = n + 1 $ one has
\[\varepsilon_{2} = -\frac{1}{(n + 2)} \varepsilon_{1}\]
and for $\nu = n - 1 $ one has
$$
\varepsilon_{2} = \frac{1}{(n - 2)} \varepsilon_{1} ,
$$
and correspondingly the functions (2.29)-(2.30) take the form
\begin{equation}
c_{n}(\tau) = {\widetilde \varepsilon}_{1}I_{n+1}(M\tau) +
{\widetilde \varepsilon}_{2}I_{n-1}(M\tau) +
\eta_{1}K_{n+1}(M\tau) +
\eta_{2}K_{n-1}(M\tau),
\end{equation}
\begin{eqnarray}
&f_{n}(\tau)& = \tau\left(-\frac{1}{(n+2)}
{\widetilde \varepsilon}_{1}I_{n+1}(M\tau)
+ \frac{1}{(n-2)}{\widetilde \varepsilon}_{2}
I_{n-1}(M\tau)\right.\nonumber \\
&&\left. -\frac{1}{(n+2)}\eta_{1}K_{n+1}(M\tau) +
\frac{1}{(n-2)}\eta_{2}K_{n-1}(M\tau)\right).
\end{eqnarray}

We have also to find the basis functions for ghosts. The eigenvalue
equations for ghosts in the de Donder gauge have the form
(see Appendix)
\[\Box\varphi_{\mu} + \lambda \varphi_{\mu} = 0\]
and the corresponding fields can be expanded on a family of
3-spheres as
\begin{equation}
\varphi_{0}(x,\tau) = \sum_{n=1}^{\infty} l_{n}(\tau) Q^{(n)}(x),
\end{equation}
\begin{equation}
\varphi_{i}(x,\tau) = \sum_{n=2}^{\infty} \biggr[
m_{n}(\tau) \frac{\nabla_{i} Q^{(n)}(x)}{(n^{2} - 1)}
+p_{n}(\tau)S_{i}^{(n)}(x)\biggr].
\end{equation}
The functions $l_{n}(\tau), m_{n}(\tau)$ and $p_{n}(\tau)$
can be found similarly to those for
harmonics of gravitational perturbations. They have the form
\begin{equation}
l_{n}(\tau) = \frac{1}{\tau}\Bigr(\kappa_{1}I_{n+1}(M\tau) +
\kappa_{2}I_{n-1}(M\tau)
+ \theta_{1}K_{n+1}(M\tau) + \theta_{2}K_{n-1}(M\tau)\Bigr),
\end{equation}
\begin{eqnarray}
&m_{n}(\tau)& = - (n-1)\kappa_{1}I_{n+1}(M\tau) +
(n+1)\kappa_{2}I_{n-1}(M\tau)\nonumber \\
&&- (n-1)\theta_{1}K_{n+1}(M\tau) +
(n+1)\theta_{2}K_{n-1}(M\tau),
\end{eqnarray}
\begin{equation}
p_{n}(\tau) = \vartheta I_{n}(M\tau) + \rho K_{n}(M\tau).
\end{equation}

\section {BOUNDARY CONDITIONS}
\hspace {\parindent}
We use the boundary conditions for linearized gravity studied
in Ref. [2]. Assuming that the spatial components $h_{ij}$ of
perturbations of the gravitational field, and the normal
component $\varphi_{0}$ of the ghost field vanish on the boundary,
the authors of Ref. [2] obtained
boundary conditions for gauge-invariant amplitudes in the form
\begin{equation}
h_{i j}|_{\partial \cal M} = h_{i 0} |_{\partial \cal M} =
\varphi_{0}|_{\partial \cal M} = 0,
\end{equation}
\begin{equation}
\left(\frac{\partial h_{00}}{\partial \tau} + \frac{6}{\tau} h_{00}
-\frac{\partial (g^{i j} h_{i j})}{\partial \tau}\right)
|_{\partial \cal M} = 0,
\end{equation}

\begin{equation}
\left(\frac{\partial \varphi_{i}}{\partial \tau} -
\frac{2}{\tau} \varphi_{i} \right)|_{\partial \cal M} = 0.
\end{equation}

Inserting the expansions (2.2)-(2.4) and (2.34)-(2.35) into
(3.1)-(3.3), one finds the following boundary
conditions on the basis functions:
\begin{equation}
\left(\frac{d a_{n}(\tau)}{d \tau} + \frac{6 a_{n}(\tau)}{\tau}
-{1\over \tau^{2}}
\frac{d e_{n}(\tau)}{d \tau}\right)|_{\partial \cal M} = 0,
\end{equation}
\begin{equation}
b_{n}(\tau)|_{\partial \cal M} = 0,
\end{equation}
\begin{equation}
c_{n}(\tau)|_{\partial \cal M} = 0,
\end{equation}
\begin{equation}
d_{n}(\tau)|_{\partial \cal M} = 0,
\end{equation}
\begin{equation}
e_{n}(\tau)|_{\partial \cal M} = 0,
\end{equation}
\begin{equation}
f_{n}(\tau)|_{\partial \cal M} = 0,
\end{equation}
\begin{equation}
k_{n}(\tau)|_{\partial \cal M} = 0,
\end{equation}
\begin{equation}
l_{n}(\tau)|_{\partial \cal M} = 0,
\end{equation}
\begin{equation}
\left(\frac{d m_{n}(\tau)}{d \tau}
- \frac{2}{\tau} m_{n}(\tau)
\right)|_{\partial \cal M} = 0,
\end{equation}
\begin{equation}
\left(\frac{d p_{n}(\tau)}{d \tau}
- \frac{2}{\tau} p_{n}(\tau)
\right)|_{\partial \cal M} = 0.
\end{equation}

Now we can calculate $\zeta(0)$ for the gravitational field on the
background which represents the part of 4-dimensional flat
Euclidean space bounded by two concentric 3-spheres
of radii $\tau_{+}$ and $\tau_{-}$ respectively.
Let us begin with the contribution of the transverse-traceless
harmonics of the gravitational field to $\zeta(0)$. On inserting
(2.13) into the boundary condition (3.10)
one gets the system of equations
\[\alpha_{1} I_{n}^{-} + \alpha_{2} K_{n}^{-} = 0,\]
\begin{equation}
\alpha_{1} I_{n}^{+} + \alpha_{2} K_{n}^{+} = 0,
\end{equation}
where
$$
I_{n}^{-} \equiv I_{n}(M \tau_{-}),
I_{n}^{+} \equiv I_{n}(M
\tau_{+}), K_{n}^{-} \equiv K_{n}(M \tau_{-}),
K_{n}^{+} \equiv K_{n}(M \tau_{+}).
$$
The condition for the existence of non-trivial
solutions of the system
(3.14) is the vanishing of its determinant, i.e.
\begin{equation}
I_{n}^{-} K_{n}^{+} - I_{n}^{+} K_{n}^{-} = 0.
\end{equation}
Eq. (3.15) provides the eigenvalue condition which we need to
apply the version of $\zeta$-function technique described
in [8-10]. It is obvious that we can neglect the first term on the
left-hand side of Eq. (3.15). Then using the power-series expansion
for $I$ and $K$ functions [9] one
can show that $I_{\rm pole}(0)$ is equal to
zero, while using the uniform asymptotic expansions
of modified Bessel functions one can show that
$I_{\rm pole}(\infty)$ vanishes as well (cf. [21]). Thus,
we only have to calculate $I_{\log}$. Looking at the uniform
asymptotic expansions for modified Bessel functions one can easily
see that the coefficient of $\log M$ in the determinant (3.15)
is equal to $(-1)$. Bearing in mind that the degeneracy for
tensor transverse-traceless harmonics is $2 (n^{2} - 4)$ [17,23],
one gets
\[
I_{\log} = \sum_{n=3}^{\infty} (n^{2} - 4) (-1) = -\zeta_{R}(-2)
+ 4 \zeta_{R}(0) + (-3) = -5,\]
where $\zeta_{R}(s)$ is the usual Riemann $\zeta$-function, and we have
used its well-known values $\zeta_{R}(-2) = 0$ and $\zeta_{R}(0) =
-1/2$. Thus, now we can write
\begin{equation}
\zeta(0)_{\rm tensor} = -5.
\end{equation}

Let us now evaluate the contribution from vector harmonics.
Inserting Eqs. (2.32)-(2.33)
into the boundary conditions (3.6) and (3.9) one has the
system of equations
\[{\widetilde \varepsilon}_{1}I_{n+1}^{-} +
{\widetilde \varepsilon}_{2}I_{n-1}^{-} +
\eta_{1}K_{n+1}^{-} +
\eta_{2}K_{n-1}^{-} = 0,\]
\[{\widetilde \varepsilon}_{1}I_{n+1}^{+} +
{\widetilde \varepsilon}_{2}I_{n-1}^{+} +
\eta_{1}K_{n+1}^{+} +
\eta_{2}K_{n-1}^{+} = 0,\]
\[-{\widetilde \varepsilon}_{1} \frac{I_{n+1}^{-}}{(n+2)}
+{\widetilde \varepsilon}_{2} \frac{I_{n-1}^{-}}{(n-2)}
-\eta_{1}\frac{K_{n+1}^{-}}{(n+2)} +
\eta_{2}\frac{K_{n-1}^{-}}{(n-2)} = 0,\]
\begin{equation}
-{\widetilde \varepsilon}_{1} \frac{I_{n+1}^{+}}{(n+2)}
+{\widetilde \varepsilon}_{2} \frac{I_{n-1}^{+}}{(n-2)}
-\eta_{1}\frac{K_{n+1}^{+}}{(n+2)} +
\eta_{2}\frac{K_{n-1}^{+}}{(n-2)} = 0.
\end{equation}

The vanishing of the determinant of the system (3.17) is the eigenvalue
condition for vector harmonics. Taking into account only the
dominant terms of this determinant (i.e. the terms including $I^{+}$
and $K^{-}$), one can write the eigenvalue condition in the form
\begin{equation}
I_{n+1}^{+} I_{n-1}^{+} K_{n+1}^{-} K_{n-1}^{-} = 0,
\end{equation}
where we have omitted the unessential common multiplier which does
not depend on $M$. One finds that just as in the case of
tensor perturbations
\[I_{\rm pole}(\infty) = I_{\rm pole}(0) = 0 \]
and we have to calculate $I_{\log}$. It is easy to see from uniform
asymptotic expansions of modified Bessel functions [24] that
\[I_{\log} = \sum_{n=3}^{\infty} (n^{2} - 1) (-2) =
-2\zeta_{R}(-2) + 2\zeta_{R}(0) + 6 = 5.\]
Here, $2(n^{2}-1)$ is the degeneracy of
vector harmonics [18,23]. Thus, we have
\begin{equation}
\zeta(0)_{\rm vector} = 5.
\end{equation}

Further to the entangled modes $c_{n}(\tau)$ and
$f_{n}(\tau)$ where $n=3,\ldots$ we have also the decoupled mode
$c_{2}(\tau)$ obeying the equation
\[\frac{d^{2}c_{2}(\tau)}{d \tau^{2}} + \frac{1}{\tau}
\frac{dc_{2}(\tau)}{d \tau} - \frac{9 c_{2}(\tau)}
{\tau^{2}} -M^{2} c_{2}(\tau) = 0,\]
whose solution can be written as
$$
c_{2}(\tau) = \varepsilon I_{3}(M\tau) + \eta K_{3}(M\tau).
$$
This function satisfies Dirichlet boundary conditions on the
3-sphere boundaries and the corresponding eigenvalue condition is
\[I_{3}^{+} K_{3}^{-} = 0.\]
Its $I_{\log}$ and correspondingly $\zeta(0)$ is
\begin{equation}
\zeta(0)_{\rm vector\,decoupled} = -3.
\end{equation}

The most complicated task is the calculation of the contribution of
scalar harmonics, since in this case we deal with the
system of four entangled equations for four basis functions
$a_{n}(\tau), b_{n}(\tau), d_{n}(\tau)$ and $e_{n}(\tau)$.
Inserting the expressions for these
functions from Eqs. (2.25)-(2.28)
into the boundary conditions (3.4), (3.5), (3.7) and (3.8)
one finds the following system of equations at each boundary:
\begin{eqnarray}
&&2\gamma_{1}(I_{n}^{\pm}- M\tau_{\pm}I_{n}'^{\pm})
+2\gamma_{2}(I_{n}^{\pm}+ M\tau_{\pm}I_{n}'^{\pm})
+\gamma_{3}(6I_{n-2}^{\pm} + 2M\tau_{\pm} I_{n-2}'^{\pm})\nonumber \\
&&+\gamma_{4}(6I_{n+2}^{\pm} + 2M\tau_{\pm} I_{n+2}'^{\pm})
+2\delta_{1}(K_{n}^{\pm}- M\tau_{\pm}K_{n}'^{\pm})
+2\delta_{2}(K_{n}^{\pm}+ M\tau_{\pm}K_{n}'^{\pm})\nonumber \\
&&+\delta_{3}(6K_{n-2}^{\pm} + 2M\tau_{\pm} K_{n-2}'^{\pm})
+\delta_{4}(6K_{n+2}^{\pm} + 2M\tau_{\pm} K_{n+2}'^{\pm}) = 0,\nonumber \\
&&\gamma_{2}I_{n}^{\pm} + (n + 1)
\gamma_{3}I_{n-2}^{\pm} - (n-1)\gamma_{4}I_{n+2}^{\pm}\nonumber\\
&&+\delta_{2}K_{n}^{\pm} + (n + 1)
\delta_{3}K_{n-2}^{\pm} - (n - 1)\delta_{4}K_{n+2}^{\pm}=0,
\nonumber\\
&&-\gamma_{2}I_{n}^{\pm} +
\frac{(n+1)}{(n-2)}\gamma_{3}I_{n-2}^{\pm}
+\frac{(n-1)}{(n+2)}\gamma_{4}I_{n+2}^{\pm}\nonumber\\
&&-\delta_{2}K_{n}^{\pm} +
\frac{(n+1)}{(n-2)}\delta_{3}K_{n-2}^{\pm}
+\frac{(n-1)}{(n+2)}\delta_{4}K_{n+2}^{\pm} = 0,\nonumber \\
&&3\gamma_{1}I_{n}^{\pm} - 2\gamma_{2}I_{n}^{\pm}
-\gamma_{3}I_{n-2}^{\pm} - \gamma_{4}I_{n+2}^{\pm}\nonumber \\
&&+3\delta_{1}K_{n}^{\pm} - 2\delta_{2}K_{n}^{\pm}
-\delta_{3}K_{n-2}^{\pm} - \delta_{4}K_{n+2}^{\pm} =0.
\end{eqnarray}

The determinant of this system of equations is rather a cumbersome
one, and we only write down its dominant part, which reads
\begin{eqnarray}
&& \left(-60 n \frac{(n^{2}-1)} {(n^{2}-4)} \; (I_{n}^{+})^{2}
\; I_{n-2}^{+} \; I_{n+2}^{+} \right. \nonumber\\
&& \left. -\frac{6 (n^{2} - 1)} {(n + 2)}
\; M\tau_{+} \; (I_{n-2}')^{+}
\; I_{n+2}^{+} \; (I_{n}^{+})^{2} \right. \nonumber\\
&& \left.-\frac{6 (n^{2} - 1)}{(n - 2)}
\; M\tau_{+} \; (I_{n+2}')^{+}
\; I_{n-2}^{+} \; (I_{n}^{+})^{2} \right)\nonumber \\
&& \times \left(-60 n \frac{(n^{2}-1)}
{(n^{2}-4)} \; (K_{n}^{-})^{2}
\; K_{n-2}^{-} \; K_{n+2}^{-} \right. \nonumber\\
&& \left. -\frac{6 (n^{2} - 1)}{(n + 2)}
\; M\tau_{-} \; (K_{n-2}')^{-}
\; K_{n+2}^{-} \; (K_{n}^{-})^{2} \right. \nonumber\\
&& \left.-\frac{6 (n^{2} - 1)}{(n - 2)}
\; M\tau_{-} \; (K_{n+2}')^{-}
\; K_{n-2}^{-} \; (K_{n}^{-})^{2} \right)
\end{eqnarray}
Inserting into Eq. (3.22) the expressions for uniform asymptotic
expansions of modified Bessel functions [24] we see that in the limit
$M \rightarrow \infty$ our determinant becomes
\[\frac{144 n^{2} (n^{2} - 1)^{2}}{(n^{2} - 4)^{2}}\]
and after taking the logarithm and expanding it
in inverse powers of $n$ we only get vanishing contributions
to $I_{\rm pole}$, and hence
\[I_{\rm pole}(\infty) = 0.\]
Using the power series for modified Bessel functions, in an analogous
way we see that in the $M \rightarrow 0$ limit
the determinant (3.22) becomes
\[-\frac{144 (n^{2} - 1)^{3}(n^{2} - 16)}
{n^{2} (n^{2} - 4)^{3}}. \]
This leads to vanishing contributions to $I_{\rm pole}$
and hence
\[I_{\rm pole}(0) = 0.\]

The calculation of $I_{\log}$ is straightforward and yields
\[I_{\log} = \sum_{n=3}^{\infty} \frac{n^{2}}{2} (-2) =
-\zeta_{R}(-2) + 5 = 5.\]
Thus, we can write
\begin{equation}
\zeta(0)_{\rm scalar} = 5.
\end{equation}

Now we evaluate the contribution of partially decoupled modes.
When $n = 2$ we have the system of equations for $a_{2}(\tau),
b_{2}(\tau)$ and $e_{2}(\tau)$, while the mode $d_{2}(\tau)$ does not
exist. Even without writing down explicitly the corresponding
determinant, one can show that its dominant part is proportional
to the product of two functions $I$, two functions $K$ and one
$I'$ multiplied by $M$ and one $K'$
multiplied by $M$ too. Hence their
contribution to $I_{\log}$, which in the case of finite number of
degrees of freedom coincides with $\zeta(0)$ (see Ref. [8]), can
be easily obtained knowing the uniform asymptotic expansions
of $I$ and $K$ and extracting from them the terms proportional to
$\log M$ in logarithms of these functions. Thus
\[I_{\log} = n^{2}/2 \times (-1)|_{n=2} = -2\]
and correspondingly
\begin{equation}
\zeta(0)_{\rm scalar\;n=2} = -2.
\end{equation}
When $n=1$ only the modes $a_{1}(\tau)$ and $e_{1}(\tau)$ survive,
and in an analogous way one can show that
\begin{equation}
\zeta(0)_{\rm scalar\;n=1} = 0.
\end{equation}

The only thing which we have to calculate now is the contribution
of ghost fields to the full $\zeta(0)$. Inserting the
functions $l_{n}(\tau), m_{n}(\tau)$ and $p_{n}(\tau)$ from Eqs.
(2.36)-(2.38) into the boundary conditions (3.11)-(3.13) we can
get the determinants for the corresponding system of equations. From
these determinants we can obtain, by the method described above, the
following contributions to $\zeta(0)$:
\begin{equation}
\zeta(0)_{\rm ghost\;vector} = \frac{1}{2},
\end{equation}
\begin{equation}
\zeta(0)_{\rm ghost\;scalar} = 0,
\end{equation}
\begin{equation}
\zeta(0)_{\rm ghost\;decoupled} = -\frac{1}{2}.
\end{equation}
Of course, the ghost contributions (3.26)-(3.28) to the full
$\zeta(0)$ should be multiplied by $-2$ [2,21], but this does
not affect our result.
Thus, by virtue of (3.16), (3.19), (3.20),
(3.23)-(3.28) the full $\zeta(0)$ value is given by
\begin{equation}
\zeta(0)_{\rm total} = 0.
\end{equation}

This result obtained by using the
$\zeta$-function technique coincides with
that obtained by using the covariant Schwinger-DeWitt technique
for manifolds with boundaries [2,19]. In fact, on using the
covariant technique on the part of
flat Euclidean space bounded by two
concentric 3-spheres we see that
the volume contribution to the $A_{2}$
Schwinger-DeWitt coefficient vanishes, while the surface
contributions from the two boundaries cancel each other.

\section {CONCLUSIONS}
\hspace {\parindent}
We have shown that, in quantum cosmology, one can obtain
unambiguous calculations of one-loop amplitudes providing
one studies flat Euclidean backgrounds bounded by two
concentric 3-spheres, and providing one takes into account
non-physical degrees of freedom and ghost modes. One then
finds that the covariant Schwinger-DeWitt technique,
and zeta-function regularization relying on a mode-by-mode
analysis of quantized fields, are in agreement.

Moreover, when the background 4-geometry has boundaries,
one finds it is no longer true that ghost modes cancel
the contribution of non-physical degrees of freedom [21].
To preserve gauge invariance one has thus to deal with
physical, non-physical and ghost modes. All these properties
have been here proved in the case of pure gravity.

In Ref. [20] it was shown that discrepancies do not occur
when massless spin-${1\over 2}$ fields are studied at one-loop
about Riemannian 4-geometries with two 3-sphere boundaries.
In Ref. [21] we have shown that, on studying vacuum Maxwell
theory within the Faddeev-Popov formalism, discrepancies in
one-loop calculations are again eliminated providing one
takes a flat Euclidean background bounded by two concentric
3-spheres. Here, we have proved that the same property
holds in the more involved case of linearized gravity.
Interestingly, in our paper the contribution
of physical degrees of freedom, i.e. transverse-traceless
gravitons, is corrected by the
contribution of non-physical degrees of freedom, while the
contribution of ghosts on the background under consideration
vanishes.

It therefore seems that, on considering Riemannian 4-geometries
with two boundaries, covariant or mode-by-mode descriptions
of quantum amplitudes are both legitimate, providing one
takes into account non-physical degrees of freedom and ghost
fields. By contrast, on studying Lorentzian 4-geometries,
the 3+1 decomposition and the extraction of physical degrees
of freedom is still valid, providing one can make sense
of the corresponding path integral. The Lorentzian regime is
in turn more relevant for the description of the time-evolution
of the physical universe studied within the framework of
the Hartle-Hawking program [25-28].

Last, but not least, we would like to emphasize that the
mode-by-mode quantization program may shed new light on
modern quantum field theory. We are currently investigating
the relation between the $I(M^{2},s)$
function defined in Eq. (1.1) [8-11],
and the zeta-function. This would enable one to relate gauge
invariance of quantum amplitudes
in the presence of boundaries to the invariance under homotopy
of the residue of a meromorphic function.
In physical language, this happens since a change of
the gauge-averaging
functional leads to a smooth variation of the corresponding
matrix of elliptic self-adjoint operators.
The residues at the origin of the meromorphic functions
occurring in this analysis may be studied by a suitable
generalization of the Atiyah-Patodi-Singer theory of
Riemannian 4-geometries with boundary [29].
Although it is unclear whether such a research program can be
completed, and applied to linearized gravity, it seems to
provide a very exciting and deep vision of gauge invariance
in quantum field theory.

\begin{center} {* * *}
\end{center}

After submitting our
paper, we became aware of Ref. [30], where the asymptotic expansion
of the heat kernel for elliptic operators
on Riemannian 4-manifolds with boundary is studied
to improve the analysis in Ref. [19]. In Ref. [31] Moss and Poletti,
using the new results in Ref. [30], have re-calculated the conformal
anomalies on Einstein spaces with boundary and, in the 1-boundary
case (i.e. the disk), they have found agreement with the results
obtained in Refs. [5,11,20] for
spin-${1\over 2}$ fields, and in Ref.
[21] for vacuum Maxwell theory.
However, it should be emphasized that the
agreement between the covariant Schwinger-DeWitt and the
non-covariant mode-by-mode calculations of the covariant Faddeev-Popov
path integral is achieved on the disk in the Lorentz gauge only.
In this gauge, the second-order differential operator for the
electromagnetic field is covariant and does not depend on the
choice of 3+1 decomposition.
For a more general class of relativistic gauges [21-22],
the mode-by-mode analysis of 1-loop quantum amplitudes is
gauge-dependent on the disk, while in
the 2-boundary case (i.e. the ring),
the $\zeta(0)$ value is gauge-independent.

As far as we can see, this last remaining discrepancy seems to
point out to serious limitations of the quantum theory when
the background 4-geometry does not admit a well-defined
3+1 decomposition. Nevertheless, in the light of the recent
literature [30-31], no conclusive argument exists which proves
that the gauge-invariance problem can only be addressed in the
2-boundary case. It also appears interesting to compare our mixed
boundary conditions for linearized gravity with the boundary
conditions studied in Ref. [32].

\newpage

\centerline {\bf ACKNOWLEDGMENTS}

\vspace {1cm}

We are indebted to Andrei Barvinsky for enlightening
correspondence. Our joint paper was supported in part by
the European Union under the Human Capital and Mobility
Program. Moreover, the research described in this publication
was made possible in part by Grant No
MAE000 from the International Science Foundation.
A. Kamenshchik is grateful to the Dipartimento
di Scienze Fisiche dell'Universit\`a
di Napoli and to the Istituto Nazionale di
Fisica Nucleare for kind hospitality and financial support
during his visit to Naples in May 1994.
The work of A. Kamenshchik was
partially supported by the Russian Foundation for
Fundamental Researches through grant No 94-02-03850-a.

\vspace {1cm}
\renewcommand{\theequation}{A.\arabic{equation}}

\centerline {\bf APPENDIX}

\vspace {1cm}

In Sec. II, the eigenvalue equations (2.5)-(2.11) are obtained
out of the $\Box$ operator, which is the elliptic operator defined
by
\begin{equation}
\Box \equiv g^{\mu \nu} \nabla_{\mu} \nabla_{\nu}.
\end{equation}
As stated in Sec. II, covariant differentiation $\nabla_{\mu}$
is performed with respect to the flat 4-metric $g$, in the spherical
local coordinates suitable for the description of flat Euclidean
4-space bounded by two concentric 3-spheres. The corresponding
perturbation of $g$ is denoted by $h$,
and $\tau$ is the Euclidean-time coordinate. Hence one finds
\begin{equation}
\Box h_{00} = {\partial^{2}h_{00} \over \partial \tau^{2}}
+{3\over \tau} {\partial h_{00}\over \partial \tau}
+{1\over \tau^{2}} h_{00 \mid i}^{\; \; \; \; \; \; \mid i}
-{4\over \tau^{3}} h_{0i}^{\; \; \; \mid i}
-{6\over \tau^{2}} h_{00}
+{2\over \tau^{2}} g^{ij}h_{ij},
\end{equation}
\begin{equation}
\Box h_{0k} = {\partial^{2} h_{0k} \over \partial \tau^{2}}
+{1\over \tau} {\partial h_{0k} \over \partial \tau}
+{1\over \tau^{2}}h_{0k \mid i}^{\; \; \; \; \; \; \mid i}
-{7\over \tau^{2}}h_{0k}
-{2\over \tau^{3}} h_{ik}^{\; \; \; \mid i}
+{2\over \tau} h_{00,k},
\end{equation}
\begin{equation}
\Box h_{ij} = {\partial^{2}h_{ij}\over \partial \tau^{2}}
-{1\over \tau} {\partial h_{ij} \over \partial \tau}
+{1\over \tau^{2}} h_{ij \mid k}^{\; \; \; \; \; \mid k}
-{2\over \tau^{2}}h_{ij}
+{2\over \tau} \Bigr(h_{i0 \mid j}+h_{j0 \mid i} \Bigr)
+{2\over \tau^{2}} \; g_{ij} \; h_{00},
\end{equation}
where a vertical stroke denotes
3-dimensional covariant differentiation
on a 3-sphere of unit radius [17-18].
In a similar way, the ghost operators acting on (2.34)-(2.35)
are found to be (cf. [21])
\begin{equation}
\Box \varphi_{0} = {\partial^{2} \varphi_{0}
\over \partial \tau^{2}}
+{3\over \tau} {\partial \varphi_{0} \over \partial \tau}
+{1\over \tau^{2}} \; \varphi_{0 \mid i}^{\; \; \; \; \mid i}
-{3\over \tau^{2}} \; \varphi_{0}
-{2\over \tau^{3}} \; \varphi_{k}^{\; \; \mid k},
\end{equation}
\begin{equation}
\Box \varphi_{i}={\partial^{2}\varphi_{i} \over \partial \tau^{2}}
+{1\over \tau} {\partial \varphi_{i} \over \partial \tau}
+{1\over \tau^{2}} \; \varphi_{i \mid k}^{\; \; \; \; \mid k}
-{2\over \tau^{2}} \; \varphi_{i}
+{2\over \tau} \; \varphi_{0,i}.
\end{equation}

\vspace {1cm}

\hrule

\vspace {1cm}

\begin{description}
\item [{[\rm 1]}]
P.A. Griffin and D.A. Kosower, Phys. Lett.
B {\bf 233}, 295 (1989).
\item [{[\rm 2]}]
I.G. Moss and S. Poletti, Nucl. Phys. B {\bf 341},
155 (1990).
\item [{[\rm 3]}]
I.G. Moss and S. Poletti, Phys. Lett. B {\bf
245}, 355 (1990).
\item [{[\rm 4]}]
S. Poletti, Phys. Lett. B {\bf 249}, 249 (1990).
\item [{[\rm 5]}]
P.D. D'Eath and G. Esposito,
Phys. Rev. D {\bf 43}, 3234 (1991).
\item [{[\rm 6]}]
P.D. D'Eath and G. Esposito,
Phys. Rev. D {\bf 44}, 1713 (1991).
\item [{[\rm 7]}]
G. Esposito, {\it Quantum Gravity, Quantum
Cosmology and Lorentzian Geometries} (Lecture Notes in Physics, m 12,
Springer-Verlag, Berlin, 1994).
\item [{[\rm 8]}]
A.O. Barvinsky, A.Yu. Kamenshchik and I.P.
Karmazin, Ann. Phys. (N. Y.) {\bf 219}, 201 (1992).
\item [{[\rm 9]}]
A.O. Barvinsky, A.Yu. Kamenshchik, I.P. Karmazin
and I.V. Mishakov, Class. Quantum Grav. {\bf 9}, L27 (1992).
\item [{[\rm 10]}]
A.Yu. Kamenshchik and I.V. Mishakov,
Int. J. Mod. Phys. A {\bf 7}, 3713 (1992).
\item [{[\rm 11]}]
A.Yu. Kamenshchik and I.V. Mishakov,
Phys. Rev. D {\bf 47}, 1380 (1993).
\item [{[\rm 12]}]
B.S. DeWitt, {\it Dynamical Theory of Groups and
Fields} (Gordon and Breach, New York, 1965).
\item [{[\rm 13]}]
J.S. Dowker and R. Critchley, Phys. Rev. D {\bf
16}, 3390 (1977).
\item [{[\rm 14]}]
S.W. Hawking, Commun. Math. Phys. {\bf 55}, 133,
(1977).
\item [{[\rm 15]}]
R. Arnowitt, S. Deser and C.W. Misner, in {\it
Gravitation: an Introduction to Current Research}, ed. by L. Witten
(Wiley, New York, 1962).
\item [{[\rm 16]}]
S.M. Christensen and M.J. Duff, Nucl. Phys. B
{\bf 170}, 480 (1980).
\item [{[\rm 17]}]
K. Schleich, Phys. Rev. D {\bf 32}, 1889 (1985).
\item [{[\rm 18]}]
J. Louko, Phys. Rev. D {\bf 38}, 478 (1988).
\item [{[\rm 19]}]
T.P. Branson and P.B. Gilkey, Commun. Part. Diff.
Eq. {\bf 15}, 245 (1990).
\item [{[\rm 20]}]
A.Yu. Kamenshchik and I.V. Mishakov,
Phys. Rev. D {\bf 49}, 816 (1994).
\item [{[\rm 21]}]
G. Esposito, A.Yu. Kamenshchik, I.V. Mishakov and
G. Pollifrone, submitted to Class. Quantum Grav.
(DSF preprint 94/4).
\item [{[\rm 22]}]
G. Esposito, Class. Quantum Grav. {\bf 11}, 905
(1994).
\item [{[\rm 23]}]
E.M. Lifshitz and I.M. Khalatnikov, Adv. Phys. {\bf
12}, 185 (1963).
\item [{[\rm 24]}]
{\it Handbook of
Mathematical Functions with Formulas, Graphs and Mathematical Tables},
edited by M. Abramowitz and I. Stegun, Natl. Bur. Stand.
Appl. Math. Ser. No 55 (U.S. GPO, Washington, D.C., 1965).
\item [{[\rm 25]}]
J. B. Hartle and S. W. Hawking, Phys. Rev. D {\bf 28}, 2960
(1983); S. W. Hawking, Nucl. Phys. B {\bf 239}, 257 (1984).
\item [{[\rm 26]}]
A.O. Barvinsky, Phys. Rep. {\bf 230}, 237 (1993).
\item [{[\rm 27]}]
A.O. Barvinsky and A.Yu. Kamenshchik, Class.
Quantum Grav. {\bf 7}, L181 (1990).
\item [{[\rm 28]}]
A.Yu. Kamenshchik, Phys. Lett. B, {\bf 316},
45 (1993).
\item [{[\rm 29]}]
M.F. Atiyah, V.K. Patodi and I.M. Singer, Math. Proc. Camb.
Phil. Soc. {\bf 79}, 71 (1976).
\item [{[\rm 30]}]
D.V. Vassilevich, {\it Vector Fields on a Disk with Mixed
Boundary Conditions} (St. Petersburg preprint SPbU-IP-94-6).
\item [{[\rm 31]}]
I.G. Moss and S. Poletti, Phys. Lett. B {\bf 333},
326 (1994).
\item [{[\rm 32]}]
A.O. Barvinsky, Phys. Lett. B {\bf 195}, 344 (1987).

\end{description}
\end{document}